\documentclass[aps,prl,reprint,superscriptaddress,preprintnumbers]{revtex4-2}
\pdfoutput=1

\usepackage[T1]{fontenc}
\usepackage[utf8]{inputenc}
\usepackage{amsmath,amssymb,amsbsy}
\usepackage{graphicx}
\usepackage{slashed}
\usepackage{booktabs}
\usepackage{xcolor}
\usepackage[colorlinks=true,citecolor=blue,urlcolor=blue,linkcolor=blue]{hyperref}
\DeclareMathAlphabet\mathbfcal{OMS}{cmsy}{b}{n}
\DeclareMathOperator{\Tr}{Tr}
\newcommand{\code}[1]{\textsc{#1}}
\newcommand{\as}{a_s}
\newcommand{\dd}{\mathrm{d}}

\begin{document}

\preprint{OUTP-26-15P, TUM-HEP-1620/26, ZU-TH 48/26}

\title{\texorpdfstring{\boldmath The three-loop QCD helicity amplitudes for $V$+jet production\\ at hadron colliders in full color}{The three-loop QCD helicity amplitudes for $V$+jet production at hadron colliders in full color}}

\author{Thomas Gehrmann}
\affiliation{Physik-Institut, Universit\"at Z\"urich, Winterthurerstrasse 190, CH-8057 Z\"urich, Switzerland}
\author{Petr Jakub\v{c}\'{i}k}
\affiliation{Rudolf Peierls Centre for Theoretical Physics, University of Oxford, Clarendon Laboratory, Parks Road, Oxford OX1 3PU, United Kingdom}
\author{Andreas von Manteuffel}
\affiliation{Institut f\"ur Theoretische Physik, Universit\"at Regensburg, 93040 Regensburg, Germany}
\author{Cesare Carlo Mella}
\affiliation{Rudolf Peierls Centre for Theoretical Physics, University of Oxford, Clarendon Laboratory, Parks Road, Oxford OX1 3PU, United Kingdom}
\author{Lorenzo Tancredi}
\affiliation{Technical University of Munich, TUM School of Natural Sciences, Physics Department, James-Franck-Stra\ss e 1, 85748 Garching, Germany}

\begin{abstract}
We compute the three-loop Quantum Chromodynamics (QCD) corrections to the production of a vector boson $V=Z,W^\pm,\gamma^*$ and a jet in
quark-antiquark annihilation and its kinematic 
crossings, retaining full dependence on the number of colors. 
By expressing the infrared-finite remainders 
in terms of a small set of rational and special functions, we find that the amplitudes live in a strongly restricted
function space, compared to the full space of the related Feynman integrals. 
We provide fast and reliable evaluation in all kinematic regions using numerical transport.

\end{abstract}
\maketitle

\textit{Introduction.} The production of an electroweak vector boson in association with a jet
is a benchmark process for precision QCD at hadron colliders. 
The clean experimental signatures of leptonic decays allow detailed studies of the boson transverse momentum and of its recoil against QCD radiation,
which probe the strong coupling and the partonic structure of the proton and provide stringent tests of perturbative predictions. 
Experimental measurements have reached a precision which matches or exceeds that of the best available next-to-next-to-leading-order (NNLO) predictions for this process~\cite{Boughezal:2015dva,Boughezal:2015ded,Gehrmann-DeRidder:2015wbt,Gehrmann-DeRidder:2017mvr,Neumann:2022lft,Alioli:2025hpa,NNLOJET:2025rno}, motivating the extension to N$^3$LO. 

Crucial ingredients for this calculation are the 
three-loop amplitudes for a vector boson coupled to a quark-
antiquark pair and a gluon. 
These amplitudes enter several complementary reactions. In decay
kinematics they describe three-jet production in $e^+e^-$
annihilation,
while kinematic crossings provide jet production in deep inelastic
scattering and vector-boson-plus-jet production at hadron colliders.
Their calculation at two loops~\cite{Garland:2002ak,Gehrmann:2011ab} was an
important ingredient of the first NNLO predictions for these processes~\cite{Gehrmann-DeRidder:2007vsv,Weinzierl:2008iv,Currie:2016ytq}.
These results were subsequently extended to axial-vector
couplings~\cite{Gehrmann:2022vuk} and to higher orders in the dimensional
regulator~\cite{Gehrmann:2023zpz}, as required for infrared subtraction at the next perturbative order.

The calculation of the master integrals entering these amplitudes has benefited from modern multiloop techniques. Building on
the differential-equation method~\cite{Kotikov:1990kg,Remiddi:1997ny,Gehrmann:1999as}
and by choice of a canonical basis~\cite{Henn:2013nsa},
the two-loop master integrals originally computed in~\cite{Gehrmann:2000zt,Gehrmann:2001ck} 
have been extended to transcendental weight
six~\cite{Gehrmann:2023etk}. Moreover, progress on the corresponding three-loop integrals~\cite{DiVita:2014pza,Canko:2021xmn,Syrrakos:2023mor,Henn:2023vbd,Gehrmann:2024tds} recently culminated in the publication of all massless three-loop four-point master integrals with one off-shell leg in a unified framework~\cite{Gehrmann:2026mln}.
In this work, two numerical evaluation strategies without a 
global polylogarithmic representation were proposed: local series expansions and numerical transport directly in physical kinematic regions.

In parallel, the closely related three-loop scattering amplitudes for Higgs-plus-jet production in the heavy-top
effective theory were obtained in~\cite{Chen:2025utl,Guan:2026igf}. Their supersymmetric counterpart, the
three-point form factor of $\Tr(\phi^2)$ has also been computed to three 
loops~\cite{Gehrmann:2024tds,Guan:2025awp}. 
On the other hand, the three-loop helicity amplitudes 
for vector-boson-plus-jet production are currently known~\cite{Gehrmann:2023jyv} only 
in the generalized leading color approximation, 
which includes only coefficients of the highest powers of the number of colors $N_c$ and the number of massless quark flavors $N_f$ and where only planar graphs contribute.
In this letter, we present the three-loop vector-boson-plus-jet
helicity amplitudes for non-singlet and vector-singlet currents in full color.

\textit{Computational setup.}
We consider the production and subsequent decay of an off-shell vector boson
\begin{equation}
  l^+(p_5)+l^-(p_6)\to V(p_4)\to \bar q(p_1)+q(p_2)+g(p_3)\,,
\label{eq:reaction}
\end{equation}
where the three external partons and all internal QCD propagators are massless.
The corresponding amplitudes for $q\bar q\to Vg$, $qg\to Vq$ and
$\bar qg\to V\bar q$ are obtained by crossing the external partons,
with the analytic continuation performed as in Ref.~\cite{Gehrmann:2002zr}.
We include the non-singlet contribution, in which the current couples to
the external quark line, and the vector-singlet contribution, in which it
couples to a closed quark loop.  
We neglect massive-quark-loop effects and
singlet axial currents which are numerically small.
Non-singlet axial couplings
can be recovered by dressing the helicity amplitudes with the appropriate
chiral electroweak couplings~\cite{Gehrmann:2023zpz,Gehrmann:2023jyv}.

Throughout, we use the dimensionless invariants
\begin{align}
  x&=\frac{2p_1\cdot p_2}{p_4^2}\,, &y&=\frac{2p_1\cdot p_3}{p_4^2}\,, &z&=\frac{2p_2\cdot p_3}{p_4^2}\,,
  \label{eq:xyz}
\end{align}
which fulfill $x+y+z=1$. 
The decay kinematic region is then defined by $x,y,z>0$.
We work in dimensional regularization with $d=4-2\epsilon$ and use the
't~Hooft--Veltman prescription for external states~\cite{tHooft:1972tcz}.

Following~\cite{Gehrmann:2023zpz}, we label the amplitudes
in decay kinematics~\eqref{eq:reaction} by
the outgoing-quark and gluon helicities and the handedness of the leptonic current.
The bare amplitude reads:
\begin{align}
 \mathcal A^V_{\lambda_{q_2}\lambda_3 \lambda_{l_5}}={}&
 -\frac{i\sqrt{4\pi\bar\alpha_s}\,(4\pi\alpha)\,C^V_{l_5l_6}}
 {D(p_4^2,m_V^2)}\,\mathbb T^a_{ij}\,
 \Big[C^V_{q_1q_2}\,M^{\rm ns}_{\lambda_{q_2}\lambda_3\lambda_{l_5}}
 \nonumber\\&
 +\delta_{q_1q_2}\sum_{f=1}^{N_f}\frac{L^V_{ff}+R^V_{ff}}{2}\,
 M^{\rm s}_{\lambda_{q_2}\lambda_3\lambda_{l_5}}\Big]\,.
 \label{eq:EWdressing}
\end{align}
Here $D(p_4^2,m_V^2)=p_4^2-m_V^2+i\Gamma_Vm_V$ and the couplings $\bar\alpha_s$ and $\alpha$ are the bare strong coupling
and the electroweak coupling.
The amplitude carries a single color structure, the fundamental
$\mathrm{SU}(N)$ generator $\mathbb T^a_{ij}$.
The terms $M^{\rm ns}$ and $M^{\rm s}$ denote the non-singlet and
vector-singlet contributions.

For a left- or right-handed
fermion current, respectively, the couplings $C_{f_1f_2}^V$ read:
\begin{align}
  L^Z_{f_1f_2}&=\frac{I_3^{f_1}-\sin^2\theta_w e_{f_1}} {\sin\theta_w\cos\theta_w}\delta_{f_1f_2}\,
 &R^Z_{f_1f_2}&=-\frac{\sin\theta_w e_{f_1}}{\cos\theta_w}
 \delta_{f_1f_2}\,,\nonumber\\
 L^\gamma_{f_1f_2}&=-e_{f_1}\delta_{f_1f_2}\,, &R^\gamma_{f_1f_2}&=-e_{f_1}\delta_{f_1f_2}\,,\nonumber\\
 L^W_{f_1f_2}&=\frac{\epsilon_{f_1,f_2}}{\sqrt2\sin\theta_w}\,,
 &R^W_{f_1f_2}&=0\,.
\end{align}
Here $\theta_w$ is the weak mixing
angle, $I_3^f=\pm1/2$ is the third component of weak isospin, and $e_f$
is the electric charge in units of $e>0$.  Furthermore,
$\epsilon_{f_1,f_2}=1$ when $f_1\ne f_2$ and they belong to the same isospin
doublet and vanishes otherwise.  Since a single charged current cannot
couple to a closed massless quark loop, the vector-singlet amplitude is
absent for $W^\pm$ production.

The amplitudes $M^{\rm ns}$ and $M^{\rm s}$ in Eq.~\eqref{eq:EWdressing}
are contractions of a QCD hadronic current with a conserved leptonic current,
retaining the full lepton-helicity and angular dependence.
For fixed partonic helicities, each can be expressed in three independent
spinor structures with scalar coefficients depending only on the partonic
invariants. We define these coefficients through the seed amplitude
for an outgoing $\bar q_R(p_1)q_L(p_2)g_+(p_3)$ state and a
left-handed leptonic current. Using
$\langle ij\rangle[ji]=2p_i\cdot p_j$, we write
\begin{align}
 M_{L+L}=\frac{[13]}{\sqrt 2 p_4^4 y z}\Big[&+
 \omega_1\langle25\rangle^2[23][56]\nonumber\\
 &+\omega_2 p_4^2\langle25\rangle[36]\nonumber\\
 &+\omega_3 \langle12\rangle[13]\langle53\rangle[36]
   \Big]\,,
 \label{eq:MLplus}
\end{align}
where $\omega_i$ are dimensionless.
The remaining helicity amplitudes follow by momentum
permutations and discrete symmetries. 
More details are provided in the Supplemental Material.

From the physical tensor decomposition of
Refs.~\cite{Gehrmann:2023zpz} we can build
projectors that extract the $\omega_i$ directly from the $Vq\bar qg$
Feynman diagrams. 
We expand each bare scalar coefficient through three loops as
\begin{equation}
  \omega_i=\sum_{\ell=0}^3 \left( \frac{\bar\alpha_s}{2\pi}\right)^\ell\omega^{(\ell)}_i
       +\mathcal O\left(\bar\alpha_s^4\right)\,.
  \label{eq:omegaexp}
\end{equation}
The spinor basis in Eq.~\eqref{eq:MLplus} isolates the Born amplitude:
\begin{align}
    \omega_1^{(0)} = 1\,\qquad\omega_2^{(0)} = 0\qquad\omega_3^{(0)} = 0\,.
    \label{eq:treelevel}
\end{align}

\begin{table}[t]
\small
\begin{tabular*}{\columnwidth}{@{\extracolsep{\fill}}cccc@{}}
\toprule
\multicolumn{4}{@{}l}{\textit{Non-singlet}, $\mathcal C_{\rm ns}$}\\
\addlinespace[2pt]
LC~\cite{Gehrmann:2023jyv} & SLC & SSLC & SSSLC\\
\midrule
$N_c^3$ & $N_c$ & $1/N_c$ & $1/N_c^3$\\
$N_fN_c^2$ & $N_f$ & $N_f/N_c^2$ & \\
$N_f^2N_c$ & $N_f^2/N_c$ &  & \\
$N_f^3$ 
&  &  & \\
\midrule
\multicolumn{4}{@{}l}{\textit{Vector singlet}, $\mathcal C_{\rm s}$}\\
\addlinespace[2pt]
$N_c^2$ & $1$ & $1/N_c^2$ & \\
$N_fN_c$ & $N_f/N_c$ &  & \\
\bottomrule
\end{tabular*}
\caption{Color factors in the three-loop non-singlet and vector-singlet
helicity amplitudes.  
Rows count closed fermion loops; columns are organized by parametric order in the color expansion.}
\label{tab:colorstructures}
\end{table}
\begin{figure}[t]
  \centering
  \includegraphics[width=0.5\textwidth]{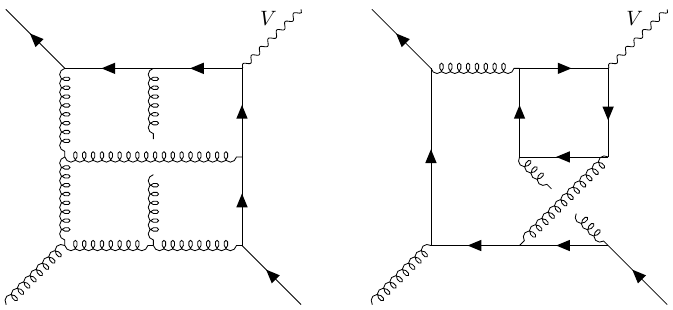}
    \caption{Representative non-planar diagrams contributing at sub-leading color to the non-singlet (left) and vector-singlet (right) parts of the $Vgq\bar{q}$ amplitude at three loops.}
  \label{fig:H_diag}
\end{figure} 

We compute all three-loop color structures for singlet and non-singlet
contributions listed in
Table~\ref{tab:colorstructures}.
In descending order of powers of the parameter $N_c\sim N_f$, 
we refer to leading color (LC) and subleading color layers.
The leading-color result of Ref.~\cite{Gehrmann:2023jyv} corresponds to the first column of the non-singlet block in Table~\ref{tab:colorstructures}, while we stress that the
$N_f^3$ term only enters through the QCD renormalization.
Figure~\ref{fig:H_diag} shows representative non-planar diagrams contributing to the subleading color layers.

We generate Feynman diagrams with
\code{QGRAF}~\cite{Nogueira:1991ex} and perform Dirac and Lorentz algebra
including the helicity projections with
\code{FORM}~\cite{Vermaseren:2000nd}. 
We map the resulting scalar integrals
onto the planar and non-planar integral families defined in~\cite{Gehrmann:2026mln}. 

We reduce them to
master integrals
using integration-by-parts identities~\cite{Tkachov:1981wb,Chetyrkin:1981qh}
and a variant of the Laporta algorithm~\cite{Laporta:2000dsw} implemented in \code{Reduze\;2}~\cite{vonManteuffel:2012np} and \code{Finred}, using 
finite-field arithmetic
and rational reconstruction~\cite{vonManteuffel:2014ixa,Peraro:2019svx}.
Instead of generating reduction tables for individual integrals, we directly reduce the amplitude.
For the rational functions, we reconstruct their dependence on $d$ and the two kinematic variables over a prime field $\mathbb F_p$.
Interestingly, sampling with a single 64-bit modulus $p$ is sufficient
to reconstruct all rational numbers in the final result.

We obtain the UV-renormalized coefficients $\omega_i^{\rm ren}$ by
expressing the bare amplitudes in terms of the renormalized
$\overline{\rm MS}$ coupling $\alpha_s(\mu)$,
documented in the Supplemental Material.
We quote results at $\mu^2=p_4^2$ for a time-like vector boson.

\textit{IR subtraction.}
We define the finite helicity coefficients by minimal subtraction of the remaining infrared poles~\cite{Catani:1998bh,Becher:2009qa,Almelid:2015jia},
\begin{equation}
 \omega_i^{\rm fin}
 =\lim_{\epsilon\to0}\mathcal Z^{-1}\omega_i^{\rm ren}.
 \label{eq:IRfactorisation}
\end{equation}
The factor $\mathcal Z=1+\mathcal O(\alpha_s)$ 
is generally a matrix in color space and obeys $\dd\log\mathcal Z/\dd\log\mu=-\Gamma$, 
where $\Gamma$ is the soft-anomalous dimension.
Since the amplitudes carry a single color
structure, in our case $\mathcal Z$ is diagonal in color space, 
and acts in the same way on all helicity coefficients
and on the non-singlet and vector-singlet contributions.
The three-loop subtraction requires the one- and two-loop amplitudes
through $\epsilon^4$ and $\epsilon^2$~\cite{Gehrmann:2023zpz}.

The soft anomalous dimension separates into
\begin{equation}
 \Gamma=\Gamma_{\rm dip}+\Delta_3.
 \label{eq:Gamma}
\end{equation}
Here $\Gamma_{\rm dip}$ describes pairwise soft color correlations and
single-parton collinear singularities. Its perturbative coefficients through three loops~\cite{Moch:2004pa,Vogt:2004mw,Moch:2005id,Moch:2005tm}
are given in the Supplemental Material.
Starting at three loops, $\Delta_3$ contains correlations involving four
color-charge operators~\cite{Almelid:2015jia}. In particular, with only three colored
external legs, no conformal cross ratio can be formed and the surviving
quadrupole correction is a kinematics-independent three-parton term.
On the $q\bar qg$ color tensor its coefficient at third order in $\alpha_s/(2\pi)$ takes the form
\begin{align}
 \Delta_3^{(3)}&=-3N_c\left(\zeta_5+2\zeta_2\zeta_3\right),
 \label{eq:Delta3}
\end{align}
where $\zeta_n$ denotes the Riemann zeta value.
Our three-loop amplitudes reproduce this non-dipole contribution.

\textit{Results.}
The renormalized and infrared-subtracted scalar coefficients $\omega_i^{\mathrm{fin}}$ are the main result of this work, attached to this letter in electronic format through three loops.
We use the canonical master integrals of~\cite{Gehrmann:2026mln}, expressed in Chen iterated integrals~\cite{Chen:1977oja} with symbol alphabet built from the following letters
\begin{equation}
\begin{alignedat}{2}
    &\text{two-loop:}\quad
        &a_i &= x_i,\quad b_i = 1-x_i \\
    &\text{parabolic:}\quad
        &P_{ij} &= x_j^2-x_j+x_i,\\
    &\text{hyperbolic:}\quad
        &H_{ij} &= 1-x_i-x_i x_j,\\
    &\text{root:}\quad
        &R_{ij} &=
        \frac{x_i x_j-\mathrm{i}\sqrt{x_i x_j x_k}}
             {x_i x_j+\mathrm{i}\sqrt{x_i x_j x_k}},
\end{alignedat}
\label{eq:alphabet}
\end{equation}
where $(x_1,x_2,x_3)=(x,y,z)$ and $i,j,k$ are distinct.
Since $d\log R_{12}+d\log R_{13}+d\log R_{23}=0$, only two root crossings are independent, giving 20 independent letters in total.
Although loop integrals acquire increasingly rich analytic structure at higher orders, the kinematics and external states strongly constrain the final amplitudes. 
This raises the question of which parts of the integral function space actually appear in the finite remainders.
A first constraint comes from the UV and IR poles of the amplitudes, which the master integrals must reproduce and which contain only the lower-loop alphabet.
The minimal transcendental basis of~\cite{Gehrmann:2026mln}, constructed to prevent cancellations between master integrals, was used to identify only 6\,232 of the 30\,770 iterated integrals containing new letters in the master-integral solutions through weight six as potential contributions to the finite part.
Our explicit calculation confirms these exclusions, and the 2\,052 observed iterated integrals suggest three additional selection criteria specifically for the decay process $V \to q\bar{q}g$:
(a) the 6 hyperbolic letters are absent, and the amplitude is described by an effective alphabet made of 14 letters; (b)
if root letters occur, the first enters as the third integration kernel; (c) the parabolic letters are never immediately followed by the ordered pairs
    \begin{align}
        P_{12}:\quad& (a_1,a_3),\ (a_1,b_2),\ (a_2,a_3),\nonumber\\
                   & (a_2,b_2),\ (a_3,a_2),\nonumber\\
        P_{21}:\quad& (a_3,a_3),\\
        P_{23}:\quad& (a_1,b_3),\ (b_3,a_3)\nonumber
    \end{align}
    and their \(y\leftrightarrow z\) images, which corresponds to $q-\bar{q}$ exchange.
Within the restricted set of~\cite{Gehrmann:2026mln}, every iterated integral containing new letters that satisfies the three criteria is present.
Interestingly, the 14-letter alphabet coincides with that of the $H$+jet amplitude~\cite{Guan:2026igf}.
We stress that after the crossing of the amplitudes into the scattering kinematics $pp \to Vj$ using the set of variables defined in~\cite{Gehrmann:2002zr}, parabolic and hyperbolic letters switch roles, but the number of
relevant letters does not change.

We express the finite remainders in 83 rational functions and 240 transcendental generators, shared by all helicity coefficients and color structures through three loops.
For color coefficient $c$ at $\ell$ loops we write
\begin{equation}
  \omega^{(\ell),c}=\sum_{r=1}^{83}R_r\,P^{(\ell),c}_r(G_1,\ldots,G_{240})\,,
  \label{eq:decomposition}
\end{equation}
where the $P_r$ are sparse polynomials with rational-number coefficients.

The generators are integrable combinations of iterated integrals and transcendental constants, each corresponding to a function of homogeneous total weight.
Some redundancy in this set is retained at lower weights 4 and 5
to avoid spurious large numerical coefficients in the corresponding
decomposition.
At weight six, only 14 independent combinations are necessary to express the complete three-loop result: three involve only the two-loop alphabet, five contain parabolic but no root letters, and six contain both root and parabolic letters.
They multiply only the four rational functions $1$, $z/y$, $z/y^2$, and $z^2/y^2$.
Moreover, through three loops, the $N_f$-dependent layers at leading color have no terms of the highest weight $2L$ at $L$ loops.
In the $N_c^L$ layer, the weight-$2L$ contribution consists of iterated integrals multiplying the rational functions in Eq.~\eqref{eq:treelevel}.
In other words, it vanishes in $\omega_2$ and $\omega_3$ and it is a pure combination of weight $2L$ in $\omega_1$.

The rational denominators contain only the six two-loop letters and two of the parabolic letters, $P_{23}$ and $P_{32}$, with individual powers no greater than four.
Besides the physical singularities at $y=0$ and $z=0$, they therefore introduce 6 apparent poles, of which the parabolic ones cross inside the bulk of the physical region. 
The parabolic denominators first appear at three loops and only in the coefficients of color factors $N_c$ and $N_f$, multiplied by weight-three combinations of iterated integrals with only two-loop letters.
The only other extended spurious singularity in the decay region is at the boundary $x\to0$.
We have checked analytically the smoothness of the scattering amplitudes across these spurious poles, which provides a consistency check of the correctness of our result.

\begin{figure}[t]
  \centering
  \includegraphics[width=0.5\textwidth]{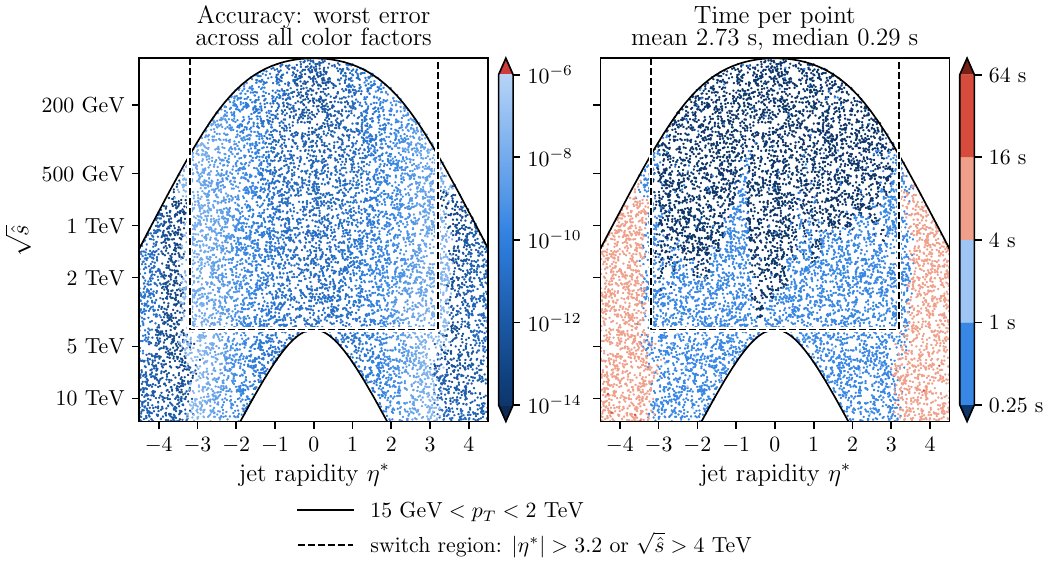}
    \caption{Numerical accuracy (left) and single-core evaluation time (right) across the sampled $V$+jet production phase space, for $\sqrt{p_4^2} = M_Z = 91.1876~\text{GeV}$. $\eta^{*}$ is the jet rapidity in the partonic frame, $p_T$ is the transverse momentum of the jet, and $\sqrt{\hat{s}}$ is the center of mass energy. The error is normalized to the largest helicity coefficient at each point; dashed lines delimit the region where two transport paths are compared.}
  \label{fig:numerics}
\end{figure}

\begin{figure*}[t]
  \centering
  \includegraphics[width=0.9\textwidth]{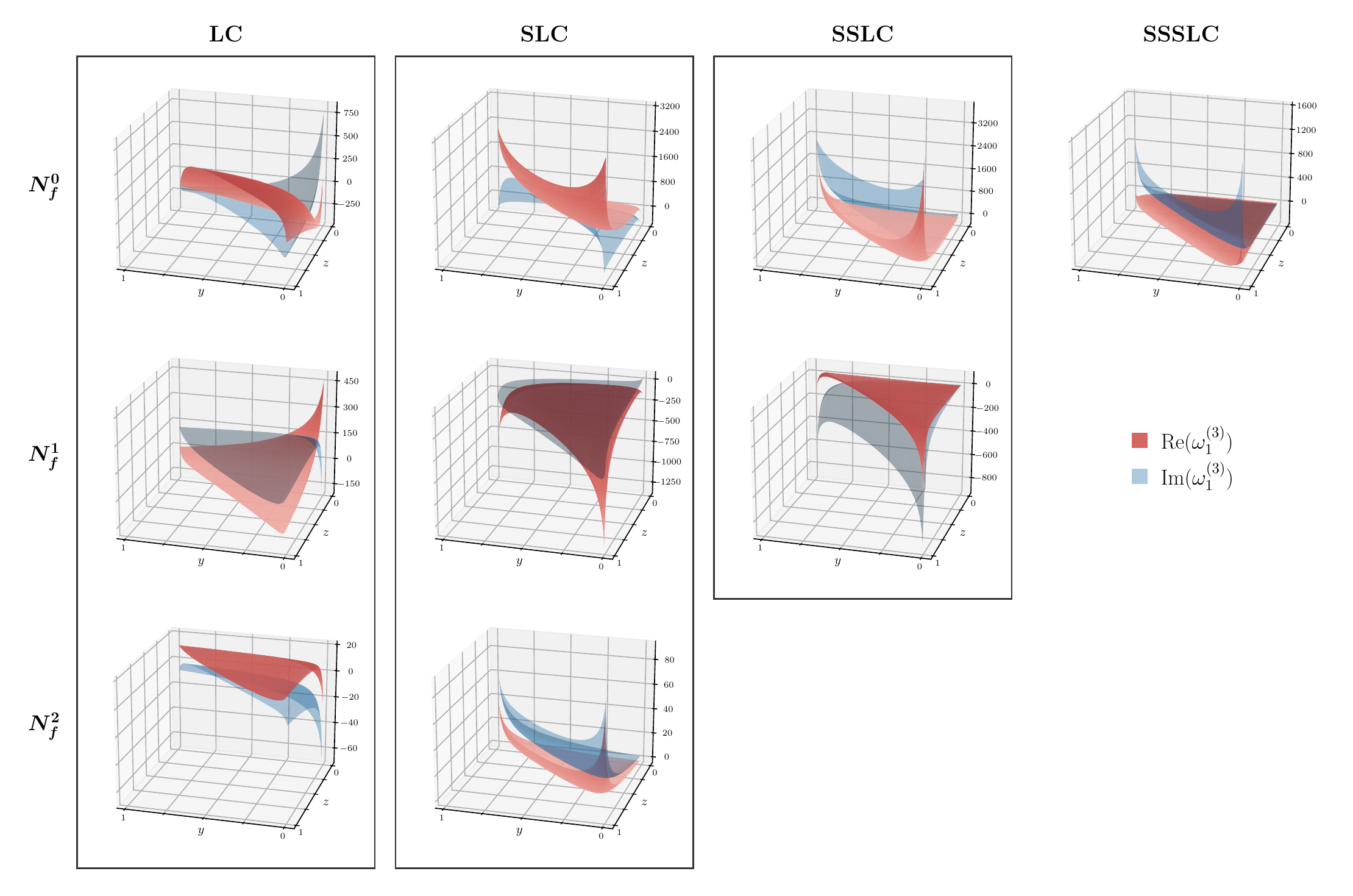}
    \caption{Real (red) and imaginary (blue) parts of the non-singlet three-loop finite remainder $\omega_1^{\mathrm{fin},(3)}$ defined in Eq.~\eqref{eq:MLplus}, plotted over the decay region $y,z>0$, $y+z<1$. Columns show the LC, SLC, SSLC, and SSSLC layers of Table~\ref{tab:colorstructures}; rows show the $N_f^0$, $N_f^1$, and $N_f^2$ contributions.}
  \label{fig:omega1-colour-layers}
\end{figure*}

\textit{Numerical evaluation.}
To evaluate the finite remainders numerically, 
we apply the method of numerical transport of~\cite{Czakon:2020vql,Czakon:2021yub,Badger:2025ljy,Czakon:2026tog}, based on the library \code{boost}~\cite{Ahnert:2011odeint}
to the individual iterated integrals as described in~\cite{Gehrmann:2026mln}. The resulting code for numerical evaluation is available in a git repository~\cite{vj3loop}.
In particular, we start from the differential equations satisfied
by the independent iterated integrals, which can be easily obtained by  direct differentiation.
To produce optimal boundary conditions for the numerical evolution in each kinematical region, we use the combination of a high-order series expansion of the solution together with one high-precision run of the evolution code~\cite{Czakon:2026tog} in double-double precision. In this way we generate one boundary condition with 27 digits in the middle of each kinematical region. 
At runtime, we can then employ the same code~\cite{Czakon:2026tog} to transport these values along complex-deformed paths to any point in each respective region. We stress that the branch of the square root $\sqrt{x y z}$ can be chosen arbitrarily since the amplitude does not depend on it.
In all physical channels, we verified agreement with an independent \code{GiNaC}~\cite{Vollinga:2004sn} evaluation of the leading-color results expressed in generalized polylogarithms~\cite{Gehrmann:2023jyv}.
Moreover, as a cross-check, we verified that the amplitudes including all
subleading-color terms satisfy the CP relation between the $qg$ 
and $\bar q g$ channels at the $10^{-10}$ level, even though the two are continued
from different corners of the phase space.

For our final numerical implementation, we evaluate points in the bulk of the phase space in double precision, and devise a strategy that dynamically determines whether to promote a point to 
quadruple precision, aiming for an error of less than $10^{-6}$ relative to the largest helicity coefficient. 
In particular, at forward rapidities and very high energies, we compare two paths with two different complex deformations 
to estimate the error and re-evaluate points with fewer than six reliable digits in double-double precision.
Similarly, whenever the target phase space point lies at a distance of less than $10^{-5}$ from a spurious singularity, we avoid numerical instabilities by an averaging procedure in its neighborhood. 

We assess the numerical precision of our implementation
in the scattering region by comparing the results
against a double-double-precision reference with tolerance $10^{-20}$ for the numerical transport at
$3\times 10^{4}$ points corresponding to a range of transverse momenta
$15~\text{GeV}<p_T<2~\text{TeV}$ and pseudorapidities $|\eta^*|<4.5$.
As shown in Fig.~\ref{fig:numerics}, with our strategy 
every tested helicity coefficient in every color 
layer meets the six-digit target.
The median cost is $0.3$~s per point on a single core.
The amplitudes are thus ready for use in any N$^3$LO calculations with resolved jets. 

The numerical evaluation also reveals regularities in the kinematic dependence of the non-singlet finite remainders, illustrated for $\omega_1^{(3)}$ in Fig.~\ref{fig:omega1-colour-layers}. Within a given order in the color expansion, the $N_f$-independent and linear-$N_f$ contributions have approximately proportional shapes with opposite signs over the bulk of the sampled phase space, leading to substantial cancellations.
Further cancellations among helicity and loop-interference contributions have the potential to reduce the relative impact of subleading-color terms in the N$^3$LO virtual contribution.

Furthermore, we have numerically verified the leading-power infrared
factorization of the three-loop finite remainder of $M_{L+L}$, separately
for each color layer. 
We approach the soft limit along fixed-ratio
trajectories, $y=\lambda r$, $z=\lambda(1-r)$, and the two collinear
limits along fixed-momentum-fraction trajectories,
$(y,z)=(t,w)$ and $(w,t)$, with fixed $0<r,w<1$ and $\lambda,t\to0$.
We compare to the convolution of the quark form
factor~\cite{Baikov:2009bg, Gehrmann:2010tu} with the soft-gluon
current~\cite{Herzog:2023sgb, Chen:2023hmk} and the helicity-dependent
$q\to qg$ or $\bar q\to\bar qg$ splitting
amplitudes~\cite{Guan:2024hlf} at this perturbative order. 
The relative discrepancy steadily
decreases at successive sampled points approaching the infrared
limits over the ranges resolved by the numerical evaluation.

\textit{Summary. }
We presented the full-color three-loop
$Vq\bar qg$ helicity amplitudes for non-singlet and vector-singlet currents.
After subtraction of infrared poles, the finite remainders are far more constrained than the underlying master integrals:
only a restricted fraction of the available transcendental structures survives,
and cancellations in their rationally weighted combinations remove apparent singularities.
This regularity yields a compact representation shared across helicities and color structures. 
High-precision numerical transport from a reference point enables reliable evaluation in all physical kinematic channels without an explicit functional representation.
Our results provide the full three-loop virtual contributions to 
future N$^3$LO studies of
$V$+jet production and crossed processes.

\textit{Acknowledgments.}
We thank Fabrizio Caola for useful discussions.
We have used AI agents for code implementation and in the editing process of this manuscript.
This work has been supported by the 
Excellence Cluster ORIGINS funded by the Deutsche
Forschungsgemeinschaft (DFG, German Research Foundation) under Germany’s Excellence Strategy – EXC-2094-390783311, by the European Research Council (ERC) under grant agreements 949279 (ERC
Starting Grant HighPHun) and 101019620 (ERC Advanced Grant TOPUP),
by the Swiss National Science Foundation (SNF) under contract 240015, by the UKRI Frontier Research Grant, underwriting the ERC Consolidator Grant precSM (UKRI946), 
and by the Science and Technology Facilities Council (STFC) under grant ST/X000761/1.

\clearpage
\onecolumngrid
\appendix
\section{UV renormalization}\label{app:UV}
We collect the conventions needed to use the helicity coefficients.
With $\as=\alpha_s/(2\pi)$, the $d$-dimensional running coupling obeys
\begin{equation}
 \frac{\dd\as}{\dd\log\mu}
 =-2\epsilon\as-2\beta_0\as^2-2\beta_1\as^3-2\beta_2\as^4
 +\mathcal O(\as^5).
\end{equation}
The bare and renormalized couplings are related by
\begin{align}
 \bar\alpha_s S_\epsilon
 &=\mu^{2\epsilon}\alpha_s(\mu)Z_{\alpha_s},
 \qquad S_\epsilon=(4\pi)^\epsilon e^{-\gamma_E\epsilon},\\
 Z_{\alpha_s}
 &=1-\as\frac{\beta_0}{\epsilon}
 +\as^2\left(\frac{\beta_0^2}{\epsilon^2}-\frac{\beta_1}{2\epsilon}\right)
 +\as^3\left(-\frac{\beta_0^3}{\epsilon^3}
 +\frac{7\beta_0\beta_1}{6\epsilon^2}-\frac{\beta_2}{3\epsilon}\right)
 +\mathcal O(\as^4)\,,
\end{align}
where $\gamma_E$ is the Euler--Mascheroni constant.
Including the overall factor $\sqrt{\bar\alpha_s}$ in the amplitude, a bare coefficient
at loop order $\ell$ carries $Z_{\alpha_s}^{\ell+1/2}$ after this
substitution.
The first three $\beta$-function coefficients are
\begin{align}
    \beta_0 &= \frac{11 C_A}{6} - \frac{2 T_F N_f}{3}\,, \\
    \beta_1 &= \frac{17 C_A^2}{6} - \frac{5 C_A T_F N_f}{3} -  C_F T_F N_f\,,\\
    \beta_2 &= -\frac{205}{72} C_A C_F N_f T_F-\frac{1415}{216} C_A^2 N_f T_F+\frac{79}{108} C_A N_f^2
   T_F^2+\frac{2857 C_A^3}{432}+\frac{11}{18} C_F N_f^2 T_F^2+\frac{1}{4} C_F^2 N_f T_F\,,
\end{align}
where the conventional QCD color factors are
\begin{equation}
    C_A = N_c,\quad C_F = \frac{N_c^2-1}{2N_c},\quad T_F = \frac{1}{2}\,.
\end{equation}

\section{IR subtraction}\label{app:IR}
Following~\cite{Becher:2009qa}, for either non-singlet or vector-singlet helicity coefficient, write
$\omega_i^{\rm ren}=\sum_{\ell\geq0}\as^\ell
\omega_i^{\rm ren,(\ell)}$ and
$\mathcal Z=1+\sum_{\ell\geq1}\as^\ell\mathcal Z^{(\ell)}$.
Expanding $\omega_i^{\rm fin}=\mathcal Z^{-1}\omega_i^{\rm ren}
=\sum_{\ell\geq0}\as^\ell\omega_i^{\rm fin,(\ell)}$ gives
\begin{equation}
\begin{aligned}
\omega_i^{\rm fin,(0)}
&= \omega_i^{\rm ren,(0)}, \\
\omega_i^{\rm fin,(1)}
&= \omega_i^{\rm ren,(1)}
- \mathbf{I}^{(1)} \omega_i^{\rm ren,(0)}, \\
\omega_i^{\rm fin,(2)}
&= \omega_i^{\rm ren,(2)}
- \mathbf{I}^{(2)} \omega_i^{\rm ren,(0)}
- \mathbf{I}^{(1)} \omega_i^{\rm ren,(1)}, \\
\omega_i^{\rm fin,(3)}
&= \omega_i^{\rm ren,(3)}
- \mathbf{I}^{(3)} \omega_i^{\rm ren,(0)}
- \mathbf{I}^{(2)} \omega_i^{\rm ren,(1)}
- \mathbf{I}^{(1)} \omega_i^{\rm ren,(2)},
\end{aligned}
\label{eq:finite_amplitudes}
\end{equation}
where $\mathbf I=\sum_{\ell\geq1}\as^\ell\mathbf I^{(\ell)}$ has coefficients
\begin{equation}
\begin{aligned}
\mathbf{I}^{(1)}
&= \mathcal{Z}^{(1)}, \\
\mathbf{I}^{(2)}
&= \mathcal{Z}^{(2)}
- \left(\mathcal{Z}^{(1)}\right)^2, \\
\mathbf{I}^{(3)}
&= \mathcal{Z}^{(3)}
- 2 \mathcal{Z}^{(2)} \mathcal{Z}^{(1)}
+ \left(\mathcal{Z}^{(1)}\right)^3.
\end{aligned}
\label{eq:subtraction_operators}
\end{equation}
For $i=K,q,g$, let
$\gamma^i=\sum_{\ell\geq0}\as^{\ell+1}\gamma^i_\ell$.
With $s_{ij}=2p_i\cdot p_j$ and
$L_{ij}=\log\!\bigl(\mu^2/(-s_{ij}-i0)\bigr)$, the scalar dipole
anomalous dimension on the $q\bar qg$ color tensor is
\begin{align}
 \Gamma_{\rm dip}&=\sum_{\ell\geq0}\as^{\ell+1}\Gamma_\ell,\\
 \Gamma_\ell&=\gamma^K_\ell\left[
 -C_F L_{12}-\frac{C_A}{2}(-L_{12}+L_{13}+L_{23})\right]
 +2\gamma^q_\ell+\gamma^g_\ell,
\end{align}
so that, at fixed coupling,
$\Gamma'_\ell=\partial\Gamma_\ell/\partial\log\mu
=-(2C_F+C_A)\gamma^K_\ell$.
In the decay region at $\mu^2=p_4^2$,
$(L_{12},L_{13},L_{23})=(-\log x,-\log y,-\log z)+i\pi(1,1,1)$.
The non-dipole term in Eq.~\eqref{eq:Gamma} begins as
$\Delta_3=\as^3\Delta_3^{(3)}+\mathcal O(\as^4)$,
with the scalar coefficient in Eq.~\eqref{eq:Delta3}.
Solving $\dd\log\mathcal Z/\dd\log\mu=-\Gamma$ gives
\begin{flalign}
\mathcal{Z}^{(1)}
&= \frac{\Gamma_0'}{4\epsilon^2}
 + \frac{\Gamma_0}{2\epsilon},&&
\label{eq:Z1}
\\[6pt]
\mathcal{Z}^{(2)}
&= \frac{\Gamma_0'^2}{32\epsilon^4}
 + \frac{\Gamma_0'}{8\epsilon^3}
   \left(\Gamma_0-\frac{3\beta_0}{2}\right)
 + \frac{1}{4\epsilon^2}
   \left(
   -\beta_0\Gamma_0
   +\frac{\Gamma_0^2}{2}
   +\frac{\Gamma_1'}{4}
   \right)
 + \frac{\Gamma_1}{4\epsilon},&&
\label{eq:Z2}
\\[6pt]
\mathcal{Z}^{(3)}
&= \frac{\Gamma_0'^3}{384\epsilon^6}
 + \frac{\Gamma_0'^2}{64\epsilon^5}
   (\Gamma_0-3\beta_0)
 + \frac{\Gamma_0'}{8\epsilon^4}
   \left(
   -\frac54\beta_0\Gamma_0
   +\frac{11}{9}\beta_0^2
   +\frac14\Gamma_0^2
   +\frac{\Gamma_1'}8
   \right)&&
\nonumber\\
&\quad
 + \frac{1}{\epsilon^3}
   \Biggl[
   \frac{\Gamma_0'}{54}
   (10C_A N_f T_F-17C_A^2+6C_F N_f T_F)
   +\Gamma_0
   \left(
   \frac{\beta_0^2}{6}
   +\frac{\Gamma_1'}{32}
   \right)
   -\frac18\beta_0\Gamma_0^2
   -\frac{5\beta_0\Gamma_1'}{72}
   +\frac{\Gamma_1\Gamma_0'}{16}
   +\frac{\Gamma_0^3}{48}
   \Biggr]&&
\nonumber\\
&\quad
 + \frac{1}{\epsilon^2}
   \Biggl[
   \Gamma_0
   \left(
   \frac{10C_A N_f T_F-17C_A^2+6C_F N_f T_F}{36}
   +\frac{\Gamma_1}{8}
   \right)
   -\frac{\beta_0\Gamma_1}{6}
   +\frac{\Gamma_2'}{36}
   \Biggr]
 + \frac{\Gamma_2+\Delta_3^{(3)}}{6\epsilon}.&&
\label{eq:Z3}
\end{flalign}

For convenience, we also reproduce the first three coefficients for the cusp anomalous dimension~\cite{Moch:2004pa,Vogt:2004mw}
\begin{flalign}
    \gamma^{K}_0 &=2&&\\
    \gamma^{K}_1 &=\left(\frac{67}{9}-\frac{\pi ^2}{3}\right) C_A-\frac{10 N_f}{9}&&\\
    \gamma^{K}_2 &=\left(-\frac{14 \zeta _3}{3}+\frac{10 \pi ^2}{27}-\frac{209}{54}\right) C_A N_f+\left(\frac{11 \zeta _3}{3}+\frac{11 \pi ^4}{90}-\frac{67 \pi ^2}{27}+\frac{245}{12}\right) C_A^2+\left(4 \zeta _3-\frac{55}{12}\right) C_F
   N_f-\frac{2 N_f^2}{27}\,,&&
\end{flalign}
the quark anomalous dimension~\cite{Moch:2005id}
\begin{flalign}
    \gamma^{q}_0 &=-\frac{3 C_F}{2}&&\\
    \gamma^{q}_1 &=\left(\frac{13 \zeta _3}{2}-\frac{11 \pi ^2}{24}-\frac{961}{216}\right) C_A C_F+\left(\frac{65}{108}+\frac{\pi ^2}{12}\right) C_F N_f+\left(-6 \zeta _3+\frac{\pi ^2}{2}-\frac{3}{8}\right) C_F^2&&\\
    \gamma^{q}_2 &=\left(-\frac{241 \zeta _3}{54}+\frac{11 \pi ^4}{360}+\frac{1297 \pi ^2}{1944}-\frac{8659}{5832}\right) C_A C_F N_f+\left(-\frac{1}{3} \pi ^2 \zeta _3-\frac{211 \zeta _3}{6}-15 \zeta _5+\frac{247 \pi
   ^4}{1080}+\frac{205 \pi ^2}{72}-\frac{151}{32}\right) C_A C_F^2\nonumber&&\\
   &+\left(-\frac{11}{18} \pi ^2 \zeta _3+\frac{1763 \zeta _3}{36}-17 \zeta _5-\frac{83 \pi ^4}{720}-\frac{7163 \pi ^2}{3888}-\frac{139345}{23328}\right)
   C_A^2 C_F+\left(\frac{32 \zeta _3}{9}-\frac{7 \pi ^4}{108}-\frac{13 \pi ^2}{72}+\frac{2953}{432}\right) C_F^2 N_f\nonumber&&\\
   &+\left(-\frac{\zeta _3}{27}-\frac{5 \pi ^2}{108}+\frac{2417}{5832}\right) C_F N_f^2+\left(\frac{2 \pi
   ^2 \zeta _3}{3}-\frac{17 \zeta _3}{2}+30 \zeta _5-\frac{\pi ^4}{5}-\frac{3 \pi ^2}{8}-\frac{29}{16}\right) C_F^3\,,&&
\end{flalign}
and the gluon anomalous dimension~\cite{Moch:2005tm}
\begin{flalign}
    \gamma^{g}_0 &=-\beta _0&&\\
    \gamma^{g}_1 &=\left(\frac{32}{27}-\frac{\pi ^2}{36}\right) C_A N_f+\left(\frac{\zeta _3}{2}+\frac{11 \pi ^2}{72}-\frac{173}{27}\right) C_A^2+\frac{C_F N_f}{2}&&\\
    \gamma^{g}_2 &=\left(-\frac{19 \zeta _3}{9}-\frac{\pi ^4}{90}-\frac{\pi ^2}{24}+\frac{1217}{216}\right) C_A C_F N_f+\left(-\frac{7 \zeta _3}{27}+\frac{5 \pi ^2}{324}-\frac{269}{11664}\right) C_A N_f^2-\frac{11}{72} C_F N_f^2-\frac{1}{8} C_F^2 N_f\nonumber&&\\
    &+\left(\frac{89 \zeta
   _3}{54}+\frac{41 \pi ^4}{1080}-\frac{599 \pi ^2}{1944}+\frac{30715}{11664}\right) C_A^2 N_f+\left(-\frac{5}{18} \pi ^2 \zeta _3+\frac{61 \zeta _3}{12}-2 \zeta _5-\frac{319 \pi ^4}{2160}+\frac{6109 \pi
   ^2}{3888}-\frac{48593}{2916}\right) C_A^3\,.&&
\end{flalign}

\section{Scale dependence of the finite remainders}\label{app:scale}
We quote the finite remainders at $\mu^2=p_4^2$. In the following,
$L=\log(\mu^2/p_4^2)$, and unprimed $\Gamma_\ell$ and unlabeled
$\omega_i^{\rm fin,(\ell)}$ are evaluated at that reference scale.
The finite remainders at a generic scale follow from
$\dd(\sqrt{\as}\,\omega_i^{\rm fin})/\dd\log\mu
=\Gamma\sqrt{\as}\,\omega_i^{\rm fin}$ with
four-dimensional running of $\as$:
\begin{alignat}{2}
\omega_i^{\rm fin,(1)}(\mu)
&=\omega_i^{\rm fin,(1)}
&&{}+\Big[\tfrac18\Gamma'_0L^2+\tfrac12(\Gamma_0+\beta_0)L\Big]
\omega_i^{\rm fin,(0)},\notag\\
\omega_i^{\rm fin,(2)}(\mu)
&=\omega_i^{\rm fin,(2)}
&&{}+\Big[\tfrac18\Gamma'_0L^2+\tfrac12(\Gamma_0+3\beta_0)L\Big]
\omega_i^{\rm fin,(1)}+\Big[\tfrac{1}{128}\Gamma_0^{\prime2}L^4
+\Gamma'_0\big(\tfrac{1}{16}\Gamma_0+\tfrac{5}{48}\beta_0\big)L^3
\notag\\
&&&{}+\big(\tfrac18\Gamma_0^2+\tfrac12\Gamma_0\beta_0
+\tfrac18\Gamma'_1+\tfrac38\beta_0^2\big)L^2
+\tfrac12(\Gamma_1+\beta_1)L\Big]\omega_i^{\rm fin,(0)},\notag\\
\omega_i^{\rm fin,(3)}(\mu)
&=\omega_i^{\rm fin,(3)}
&&{}+\Big[\tfrac18\Gamma'_0L^2+\tfrac12(\Gamma_0+5\beta_0)L\Big]
\omega_i^{\rm fin,(2)}+\Big[\tfrac{1}{128}\Gamma_0^{\prime2}L^4
+\Gamma'_0\big(\tfrac{1}{16}\Gamma_0+\tfrac{11}{48}\beta_0\big)L^3
\notag\\
&&&{}+\big(\tfrac18\Gamma_0^2+\Gamma_0\beta_0
+\tfrac18\Gamma'_1+\tfrac{15}{8}\beta_0^2\big)L^2
+\tfrac12(\Gamma_1+3\beta_1)L\Big]\omega_i^{\rm fin,(1)}\notag\\
&&&{}+\Big[\tfrac{1}{3072}\Gamma_0^{\prime3}L^6
+\tfrac{1}{768}\Gamma_0^{\prime2}(3\Gamma_0+7\beta_0)L^5
+\tfrac{\Gamma'_0}{192}
\big(3\Gamma_0^2+16\Gamma_0\beta_0+3\Gamma'_1
+17\beta_0^2\big)L^4\notag\\
&&&{}+\tfrac{L^3}{48}\big(\Gamma_0^3
+9\Gamma_0^2\beta_0+23\Gamma_0\beta_0^2+15\beta_0^3
+3\Gamma_0\Gamma'_1+3\Gamma_1\Gamma'_0
+5\Gamma'_0\beta_1+7\Gamma'_1\beta_0\big)\notag\\
&&&{}+\tfrac{L^2}{8}\big(2\Gamma_0\Gamma_1
+4\Gamma_0\beta_1+6\Gamma_1\beta_0+8\beta_0\beta_1
+\Gamma'_2\big)
+\tfrac L2\big(\Gamma_2+\Delta_3^{(3)}+\beta_2\big)
\Big]\omega_i^{\rm fin,(0)}.
\label{eq:scaledep}
\end{alignat}

\section{Projectors and discrete symmetries}
At $p_4^2=1$, the projectors onto the seed-helicity coefficients
$\omega_i$ use $\mathcal D_\omega=8(d-3)y^2x^2$ and
$Q=d(y-xz)-2y(1+x)$:
\begin{flalign}
\mathcal P_{\omega_1}
&=\frac{1}{\mathcal D_\omega}\Bigl[
 -\bigl((d-4)yx+dz\bigr)T_1^\dagger
 +Q T_2^\dagger-yx(y+z)T_3^\dagger
 -y(1-y)(y+z)T_4^\dagger
 -y(y-xz)T_5^\dagger
 +yz(y+z)T_6^\dagger\Bigr],&&
\\[6pt]
\mathcal P_{\omega_2}
&=\frac{1}{\mathcal D_\omega}\Bigl[
 (1-y)\bigl((8-d)yx-dz\bigr)T_1^\dagger
 +\bigl(4yx^2+(1-y)Q\bigr)T_2^\dagger
 +yx(xy-z)T_3^\dagger
 -y(1-y)^2(y+z)T_4^\dagger\notag\\
&\quad
 -y(1-y)(y-xz)T_5^\dagger
 +yz(z-xy)T_6^\dagger\Bigr],&&
\\[6pt]
\mathcal P_{\omega_3}
&=\frac{1}{\mathcal D_\omega}\Bigl[
 -d(1-y)^2T_1^\dagger
 +\bigl(2y(x-z)+d(yz-x)\bigr)T_2^\dagger
 -y(1-y)xT_3^\dagger
 +y(1-y)^2T_4^\dagger
 +y(x-yz)T_5^\dagger\notag\\
&\quad+y(1-y)zT_6^\dagger\Bigr].&&
\end{flalign}
Here $T_i^\dagger$ are the adjoints of the following open Dirac tensors,
with $\mu$ the gluon index and $\nu$ the vector-boson index:
\begin{align}
T_1^{\mu\nu}&=\slashed p_3\,p_1^\mu p_1^\nu,&
T_2^{\mu\nu}&=\slashed p_3\,p_1^\mu p_2^\nu,&
T_3^{\mu\nu}&=\slashed p_3\,g^{\mu\nu},\nonumber\\
T_4^{\mu\nu}&=\gamma^\mu p_1^\nu,&
T_5^{\mu\nu}&=\gamma^\mu p_2^\nu,&
T_6^{\mu\nu}&=\gamma^\nu p_1^\mu.
\end{align}
This is the parity-even basis of Ref.~\cite{Gehrmann:2022vuk} with its
Lorentz-index order reversed. The projectors use $p_2$ as the gluon
reference momentum; the dagger acts on the open Dirac string in the
projector contraction. For the same seed helicity, the coefficients
$\alpha_i$ of Ref.~\cite{Gehrmann:2023zpz} and ours are related by
\begin{equation}
 \omega_1=\frac{yz}{4}\alpha_2,\qquad
 \omega_2=\frac{yz}{4}\bigl[\alpha_3+(1-y)\alpha_2\bigr],\qquad
 \omega_3=\frac{yz}{4}\alpha_1.
\end{equation}
All eight helicity configurations can then be obtained as follows. We denote by $M$ the helicity configuration of Eq.~\eqref{eq:MLplus}. Let the operator $\mathsf \sigma_{ij}$ globally swap particles $i$ and $j$. The operator $\mathsf P_{\mathrm{sp}}$ maps $\langle ij\rangle\mapsto[ji]$ and $[ij]\mapsto\langle ji\rangle$ without conjugating the scalar coefficients.
They act on the complete amplitude, including all spinor products,
Mandelstam invariants, and the arguments of the form factors.
Then
\[
\boxed{
\begin{array}{c|cc}
(q,g)
& e^+_Le^-_R
& e^+_Re^-_L
\\ \hline
L,+&
 M&
\mathsf \sigma_{56} M
\\[1mm]
L,-&
-\mathsf P_{\mathrm{sp}}\mathsf \sigma_{12}\mathsf \sigma_{56} M&
-\mathsf P_{\mathrm{sp}}\mathsf \sigma_{12} M
\\[1mm]
R,-&
\mathsf P_{\mathrm{sp}}\mathsf  \sigma_{56} M&
\mathsf P_{\mathrm{sp}} M
\\[1mm]
R,+&
-\mathsf \sigma_{12} M&
-\mathsf \sigma_{12}\mathsf \sigma_{56} M
\end{array}
}
\]
where every operator acts on the complete spinor-helicity
expression to its right.

\clearpage
\twocolumngrid
\bibliography{main}
\end{document}